\documentclass[aip,amsmath,amssymb,reprint]{revtex4-1}

\usepackage{graphicx}% Include figure files
\usepackage{dcolumn}% Align table columns on decimal point
\usepackage{bm}% bold math
\usepackage[utf8]{inputenc}
\usepackage[T1]{fontenc}
\usepackage{mathptmx}
\usepackage{etoolbox}
\makeatletter
\def\@email#1#2{%
 \endgroup
 \patchcmd{\titleblock@produce}
  {\frontmatter@RRAPformat}
  {\frontmatter@RRAPformat{\produce@RRAP{*#1\href{mailto:#2}{#2}}}\frontmatter@RRAPformat}
  {}{}
}%
\makeatother

\begin{document}
\title{Triple-top-gate technique for studying the strongly-interacting 2D electron systems in heterostructures}
\author{M.~Yu.\ Melnikov}
\affiliation{Institute of Solid State Physics, Chernogolovka, Moscow District 142432, Russia}
\author{A.~A.\ Shashkin}
\affiliation{Institute of Solid State Physics, Chernogolovka, Moscow District 142432, Russia}
\author{S.-H.\ Huang}
\affiliation{Department of Electrical Engineering and Graduate Institute of Electronics Engineering, National Taiwan University, Taipei 106, Taiwan}
\author{C.~W.\ Liu}
\affiliation{Department of Electrical Engineering and Graduate Institute of Electronics Engineering, National Taiwan University, Taipei 106, Taiwan}
\author{S.~V.\ Kravchenko}\email{s.kravchenko@northeastern.edu}
\affiliation{Physics Department, Northeastern University, Boston, Massachusetts 02115, USA}
\date{\today}

\begin{abstract}
We have developed a technique that dramatically reduces the contact resistances and depletes a shunting channel between the contacts outside the Hall bar in ultra-high mobility SiGe/Si/SiGe heterostructures. It involves the creation of three overlapping independent gates deposited on top of the structure and allows transport measurements to be performed at millikelvin temperatures in the strongly interacting limit at low electron densities, where the energy of the electron-electron interactions dominates all other energy scales. This design allows one to observe the two-threshold voltage-current characteristics that are a signature for the collective depinning and sliding of the electron solid.
\end{abstract}

\maketitle

The ultra-high mobility Si/SiGe heterostructures have proven a handy tool for studying the effects of strong particle-particle correlations on the transport and thermodynamic properties of two-dimensional (2D) carrier systems.  In particular, the fractional quantum Hall effect (see, \textit{e.g}., Refs.~\cite{nelson1992observation,lai2004fractional,lai2004two,dolgopolov2021valley}), the 2D metal-insulator transition \cite{coleridge1997metal,lam1997scaling,melnikov2019quantum}, the band flattening \cite{melnikov2017indication,dolgopolov2022band}, the metallic state in a spinless two-valley electron system \cite{melnikov2020metallic}, strong enhancement of the effective mass and its independence of the electrons' spins \cite{melnikov2023spin}, strong correlations and spin effects in transport \cite{shashkin2020manifestation,shashkin2021metal,shashkin2022spin}, and collective depinning and sliding of the quantum Wigner solid \cite{melnikov2024collective} have been observed in this 2D carrier system.  To study the strongly correlated regime, one must reach the limit where the Coulomb energy, $E_{\text C}$, greatly exceeds the Fermi energy, $E_{\text F}$.  The strength of electron interactions is characterized by the interaction parameter $r_s$, defined as the ratio between the Coulomb and Fermi energies and equal to $r_{\text{s}}=g_{\text{v}}/(\pi n_{\text{s}})^{1/2}a_{\text{B}}$, where $g_{\text{v}}$ is the valley degeneracy, $n_{\text s}$ is the areal density of the electrons, and $a_{\text{B}}$ is the effective Bohr radius in a semiconductor.  Therefore, the lower the electron density, the stronger the interactions.  However, at low electron densities, the resistance of the contacts to the 2D layer of the electrons for Si-based systems dramatically increases at millikelvin temperatures, which makes the transport measurements extremely challenging. To overcome this obstacle, a so-called split-gate technique was developed \cite{kruithof1991temperature,wang1992quantum,heemskerk1998nonlinear}.  Gaps in the gate metallization were introduced, which split the gate into several parts to which the voltage can be applied to maintain a high electron density near the doped contact regions, whereas, in the main part of the sample, the electron density can be controlled independently.  These gaps were narrow enough ($<100$~nm) for the given gate-oxide thickness to provide a smoothly descending electrostatic potential from the high-density part to the low-density part. Note that the split-gate technique was applied for nanowires as well \cite{ciriano2021spin}.  Alternatively, independent control of the electron density in different parts of the sample can, in principle, be reached using a technique of overlapping gates (see, \textit{e.g.},
Refs.~\cite{angus2007gate,lim2009observation,vanbeveren2010overlapping,mittag2018edgeless,mittag2019Ggate,iwakiri2022gate,fogarty2018integrated}, where the overlapping gates separated by Al oxide as an insulator were used).

In this Letter, we report a triple-top-gate technique that dramatically reduces the contact resistances and depletes the shunting channel between the contacts outside the Hall bar in ultra-high mobility SiGe/Si/SiGe heterostructures.  This allows transport measurements to be performed at millikelvin temperatures and extremely low electron densities ($r_{\text{s}}>20$).  Also, we use SiO, rather than Al oxide, to separate the gates, which can be done in a simple evaporator and allows one to reach the highest mobilities in this electron system.  In contrast to the planar design of split gates used previously, we use a vertical design where different gates do not lie in the same plane and overlap, in which case, for the parallel-plate capacitor, the electron density in a part of the sample is controlled by the gate closest to this part.  We measure voltage-current ($V$-$I$) characteristics in both double-gate and triple-gate samples.  In the double-gate samples, as the electron density is decreased, the single-threshold $V$-$I$ characteristics stop changing below a particular value of $n_{\text s}$, which indicates the presence of a shunting conduction channel. Therefore, the strongly interacting limit at low electron densities cannot be reached in such samples.  In triple-gate samples, the shunting channel can be depleted, which allows access to the low electron densities and observation of the two-threshold $V$-$I$ characteristics that are a signature for the collective depinning and sliding of the electron solid, as has been established in Refs.~\cite{melnikov2024collective,brussarski2018transport}.

\begin{figure}
\includegraphics[width=9cm]{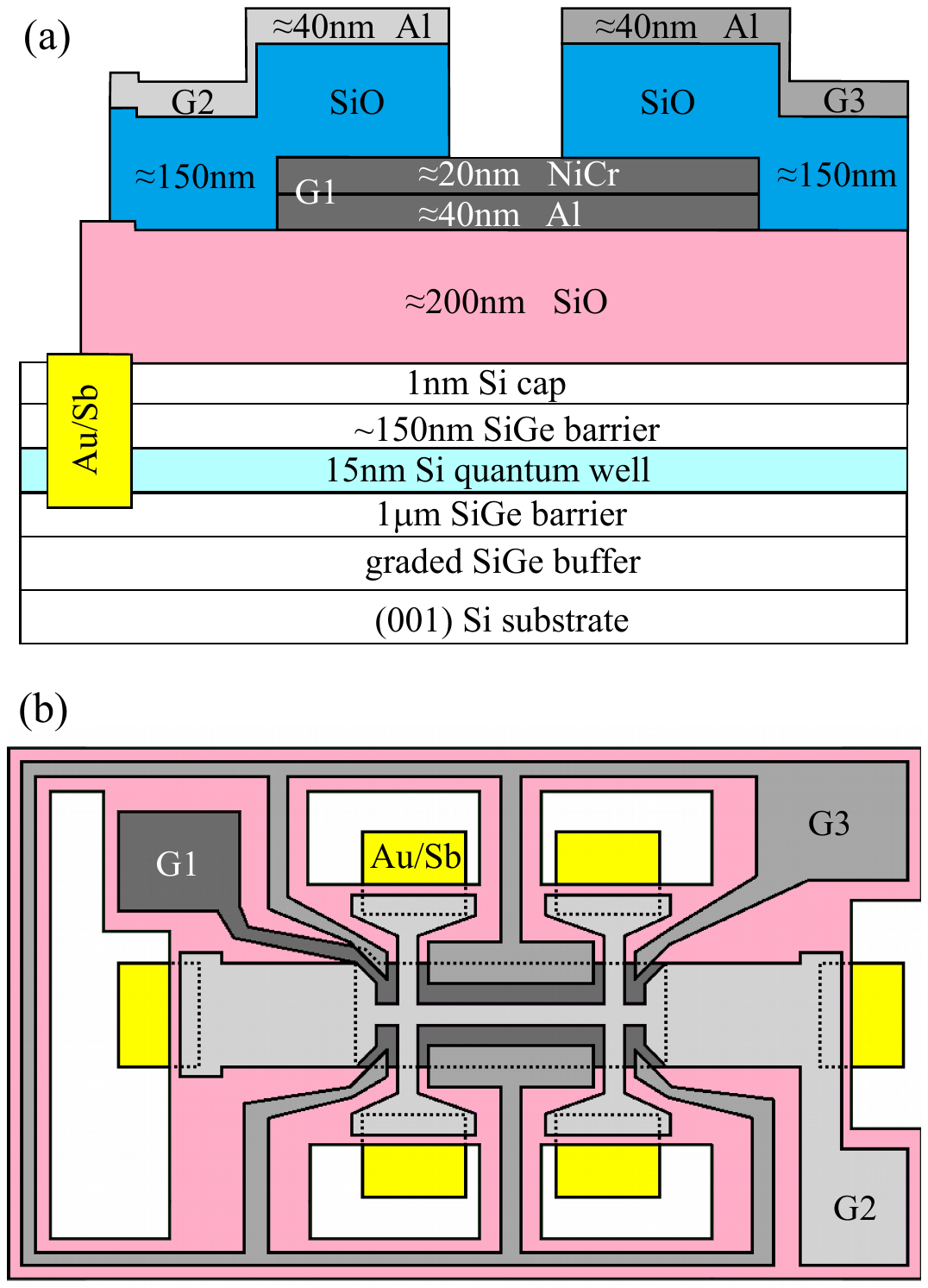}
\caption{\label{fig:scheme} (a) Layer growth sequence and the cross-section of the triple-gate samples. The Hall-bar-shaped gate G1 is located under contact gate G2 and additional gate G3. The layer thicknesses in the double-gate structures are different in that the first SiO deposition is 80~nm thick, the second SiO deposition is 200~nm thick, and the second Al gate is 60~nm thick. (b) Schematic top view on the triple-gate samples. The Hall-bar-shaped gate G1 is shown by the dashed line in the central part of the sample.}
\end{figure}

To create a sample, we utilized ultra-clean UHCVD-grown SiGe/Si/SiGe quantum wells (for details, see Refs.~\cite{lu2009observation,lu2010erratum,huang2012mobility}).  The electron mobility in our samples at low temperatures reached approximately 200~m$^2$/Vs, which is higher than that in previously used SiGe-based structures \cite{schaffler1997high,lu2009observation,lu2010erratum,huang2012mobility}.  Silicon (001) quantum well, approximately 15~nm wide, was sandwiched between Si$_{0.8}$Ge$_{0.2}$ potential barriers (Fig.~\ref{fig:scheme}). Contacts to the 2D layer were made using approximately 300~nm thick Au$_{0.99}$Sb$_{0.01}$ alloy deposited in a thermal evaporator, followed by annealing for 3-5 minutes at 440$^\omicron$~C in an N$_2$ atmosphere.  Using photo-lithography, the samples were patterned into Hall-bar shapes with a 150~$\mu$m distance between the potential probes and a 50~$\mu$m width.  An approximately 80~nm thick SiO layer was deposited on the wafer surface using a thermal evaporator, followed by the deposition of an approximately 40~nm thick Al gate G1 on top of the SiO.  Furthermore, an approximately 20~nm thick layer of NiCr was deposited on top of Al to improve the adhesion of subsequent layers.  Following this, the contact gate G2 was fabricated. The structure was covered by an approximately 200~nm thick SiO layer, and an approximately 60~nm thick aluminum gate was deposited on top of the SiO.  The contact gate was used to maintain a high electron density of approximately $2\times10^{11}$~cm$^{-2}$ near the contact areas, independent of the electron density in the main part of the sample.  No intentional doping was used, and the electron density was controlled by applying a positive DC voltage to the gate in relation to the contacts.  We applied saturating infrared illumination (which leads to the creation and diffusion in the wafer of photoexcited electrons and holes) for several minutes to the samples, after which the quality of contacts enhanced and the electron mobility increased, as had been found empirically \cite{melnikov2015ultra,melnikov2017unusual}.  The resistance of the contacts was below 10~kOhm.

In addition to the double-gate samples described above, we also studied triple-gate samples. The purpose of the third gate was to deplete the shunting channel between the contacts outside the Hall bar.  This shunting channel can manifest itself at the lowest electron densities in the insulating regime.  Triple-gate samples were patterned in Hall-bar shapes with a 50~$\mu$m width and 100~$\mu$m distance between the potential probes.  The main NiCr/Al Hall-bar gate G1 was made similarly, with the difference that an approximately 200~nm thick SiO layer was evaporated.  Furthermore, we used a two-resist undercut technique to obtain smooth edges of the evaporated films and enhance the outcome of usable samples.  The contact gate G2 was fabricated by covering the structure with an approximately 150~nm thick SiO layer, and then depositing an approximately 40~nm thick aluminum gate on top of the SiO.  An extra aluminum gate G3 to deplete the shunting channel was fabricated simultaneously with the contact gate.

We performed measurements in an Oxford TLM-400 dilution refrigerator.  The investigation primarily focused on measuring the voltage-current characteristics in the insulating regime. This involved applying a DC voltage between the source and the nearest potential probe over a distance of 25~$\mu$m; the current was measured by a transimpedance amplifier connected to a digital voltmeter.  The voltage-current curves were a little asymmetric with respect to the reversal of the voltage; for convenience, we plotted only the negative part versus the absolute value of voltage here.  The electron density was determined utilizing a standard four-terminal lock-in technique for the measurement of Shubnikov-de~Haas oscillations within the metallic regime. Measurements were conducted within a frequency range of 1-11~Hz, while maintaining currents of 0.04-4~nA within the linear response regime.  Two double-gate samples and four triple-gate samples were studied; the results obtained on all triple-gate samples were similar.

\begin{figure}
\includegraphics[width=6.7cm]{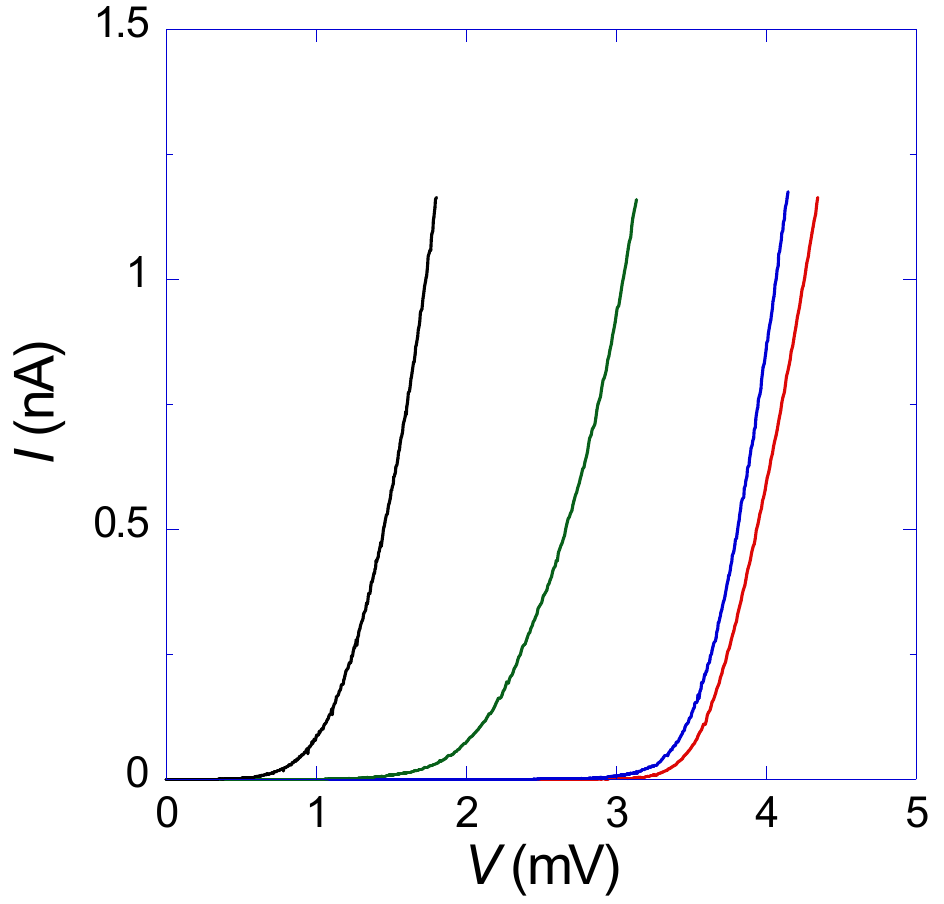}
\caption{\label{fig:2gates} $V$-$I$ characteristics measured in double-gate samples at different electron densities (from left to right): 7.95, 6.98, 6.21, and 5.82$\times10^9$~cm$^{-2}$.  $T=30$~mK.}
\end{figure}

In Fig.~\ref{fig:2gates}, we show the $V$-$I$ characteristics measured in the insulating regime at low electron densities ($r_{\text{s}}>20$) in double-gate samples.  With increasing applied voltage, the current remains near zero up to a certain threshold voltage; beyond this threshold, the current sharply increases, the threshold voltage increasing as the electron density decreases.  However, these single-threshold $V$-$I$ characteristics stop changing below $n_{\text s}\approx6\times10^9$~cm$^{-2}$. This indicates the presence of a shunting conduction channel outside the Hall bar that is obviously related to residual unintentional donor impurities in the SiGe/Si/SiGe heterostructures. Thus, the strongly interacting limit at low electron densities cannot be reached in such samples.

\begin{figure}
\includegraphics[width=6.7cm]{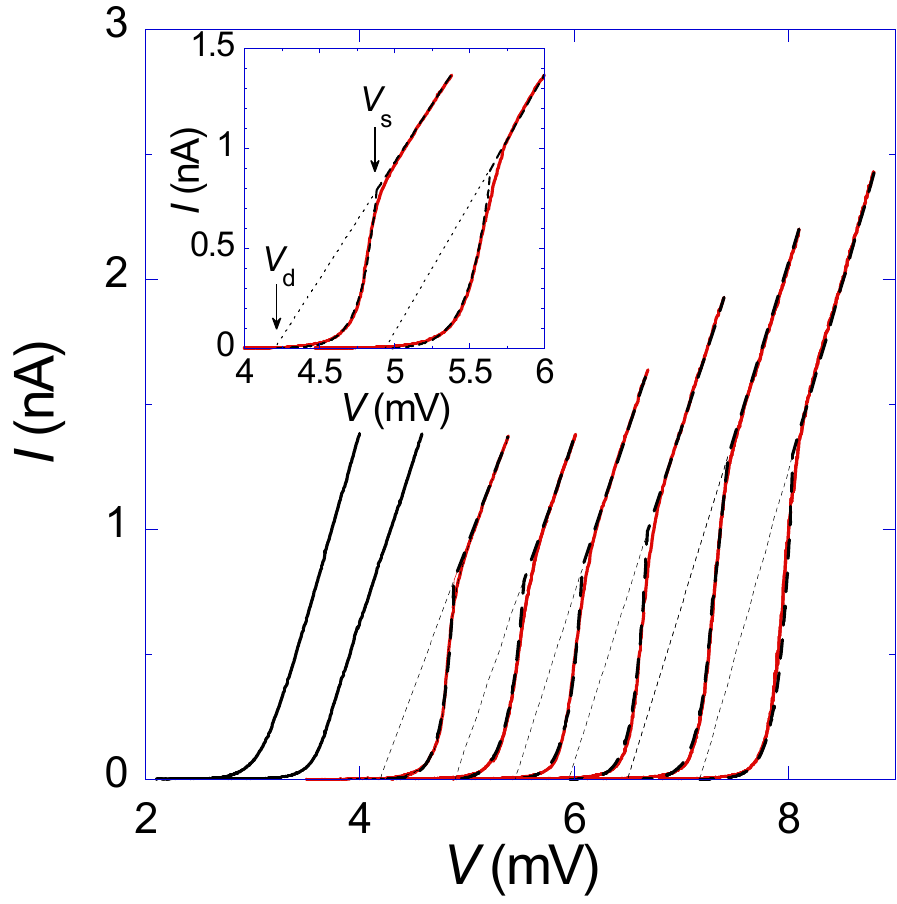}
\caption{\label{fig:3gates} $V$-$I$ characteristics measured in triple-gate samples at different electron densities (from left to right): 6.37, 6.19, 6.01, 5.92, 5.83, 5.74, 5.65, and 5.56$\times10^9$~cm$^{-2}$. $T=30$~mK.  The dashed lines are fits to the data; see text.  The inset shows $V$-$I$ characteristics for $n_{\text s}=6.01\times10^9$~cm$^{-2}$, $5.92\times10^9$~cm$^{-2}$ on an expanded scale.  Arrows indicate the dynamic threshold $V_{\text d}$, obtained by the extrapolation (dotted line) of the linear part of the $V$-$I$ curves to zero current, and the static threshold $V_{\text s}$.}
\end{figure}

Applying a negative voltage to the additional gate in triple-gate samples depletes the shunting channel.  However, we do not reach the total depletion to be in the parallel-plate capacitor regime in which the additional gate does not influence the transport properties of the 2D electron system in the main part of the sample. The suppression of the shunting channel reveals quite a different behavior of the $V$-$I$ characteristics with decreasing electron density, as shown in Fig.~\ref{fig:3gates}. The observed $V$-$I$ characteristics change with decreasing $n_{\text s}$ in the entire range of electron densities studied. At the lowest electron densities, a second threshold voltage arises on the $V$-$I$ curves. Namely, with increasing applied voltage, the current stays near zero up to the first threshold voltage, and then the current sharply increases until the second threshold voltage is reached.  Once the second threshold voltage is exceeded, the slope of the $V$-$I$ curves decreases, and the behavior becomes linear but not ohmic. These characteristics are similar to those observed earlier in silicon metal-oxide-semiconductor field-effect transistors in Ref.~\cite{brussarski2018transport}. This signals that the strongly interacting limit at low electron densities can be reached in the 2D electron system in SiGe/Si/SiGe heterostructures (for more details, see Ref.~\cite{melnikov2024collective}).

Within the phenomenological theory of pinned elastic structures, the observed two-threshold $V$-$I$ characteristics are seen as evidence for the collective depinning of an electron solid.  The model of the collective depinning of the vortex lattice in type-II superconductors \cite{blatter1994vortices} was adapted for the case of the electron solid in recent studies \cite{brussarski2018transport,melnikov2024collective}.  As the applied voltage increases, the depinning of the electron solid is indicated by the appearance of a current.  Between the dynamic ($V_{\text d}$) and static ($V_{\text s}$) thresholds, the collective pinning of the electron solid occurs, and the transport is thermally activated: $I=\sigma_0\,(V-V_{\rm d})\,\exp[-U_{\rm c}(1-V/V_{\rm s})/k_{\rm B}T]$, where $U_{\rm c}$ is the maximal activation energy of the pinning centers, $\sigma_0$ is a coefficient, and $V_{\rm d}$ corresponds to the pinning force.  When the voltage exceeds the static threshold, the electron solid slides with friction over a pinning barrier, as determined by the balance of the electric, pinning, and friction forces, resulting in linear $V$-$I$ characteristics.  The corresponding fits, shown by the dashed lines in Fig.~\ref{fig:3gates}, describe the experimental two-threshold $V$-$I$ characteristics well (for more detailed studies of the activation behavior, see Refs.~\cite{brussarski2018transport,melnikov2024collective}).

Following the definition of $r_{\text{s}}$, we see some evidence for the behavior of the electron solid. Our fabricated structure should allow further detailed measurements.

In summary, we have developed a triple-top-gate technique that dramatically reduces the contact resistances and excludes the influence of the shunting channel outside the Hall bar in ultra-high mobility SiGe/Si/SiGe heterostructures. Our vertical design with three overlapping independent gates allows transport measurements to be performed at millikelvin temperatures in the strongly interacting limit at low electron densities, where the energy of the electron-electron interactions dominates all other energy scales. In particular, this design allows one to observe the two-threshold $V$-$I$ characteristics that are a signature for the collective depinning and sliding of the electron solid.

\begin{acknowledgments}
This investigation was supported by RSF (project no. 24-22-00122).
\end{acknowledgments}

\section*{Data Availability Statement}
The data that support the findings of this study are available from the corresponding author upon reasonable request.

%\bibliography{references}
%\end{document}


\end{document}